\documentclass[runningheads]{llncs}
\usepackage{graphicx}
\usepackage{amsmath}
\usepackage{xcolor}
\usepackage{xspace}
\usepackage{comment}
\newcommand{\msgemm}{msGeMM\xspace}

\begin{document}
\title{Look-Up mAI GeMM: Increasing AI GeMMs Performance by Nearly $2.5\times$ via \msgemm}
\titlerunning{Look-Up mAI GeMM}
\author{Saeed Maleki}
\authorrunning{Saeed Maleki}
\institute{Microsoft Research \\
\email{saemal@microsoft.com}\\
}
\maketitle              % typeset the header of the contribution
\begin{abstract}
AI models are increasing in size and recent advancement in the community has shown that unlike HPC applications where double precision datatype are required, lower-precision datatypes such as fp8 or int4 are sufficient to bring the same model quality both for training and inference. Following these trends, GPU vendors such as NVIDIA and AMD have added hardware support for fp16, fp8 and int8 GeMM operations with an exceptional performance via Tensor Cores. However, this paper proposes a new algorithm called \msgemm which shows that AI models with low-precision datatypes can run with $\approx 2.5\times$ fewer multiplication and add instructions. Efficient implementation of this algorithm requires special CUDA cores with the ability to add elements from a small look-up table at the rate of Tensor Cores.

\end{abstract}
\section{Introduction}
Artificial Intelligent (AI) recent advancements have shown remarkable results for Large Language Models (LLM) such as GPT-4~\cite{gpt4} or Llama-2~\cite{llama2}. However, such models require very large number of parameters to store the weights. Luckily, recent works have shown that unlike HPC applications with fp64 datatypes, low-precision datatypes such as fp8~\cite{fp8} or int4~\cite{int4} are as effective as their wider counterparts. This allows to lower the memory requirement per parameter and have much faster compute capabilities in GPUs. As the datatypes are getting extremely narrow, interesting alternative approaches could be utilized to calculate an output of a model.

General Matrix Multiply (GeMM) is the basic operation for any AI model. Model weights and activation vectors for different input samples (stacked as a matrix) make up the two matrices for a GeMM operation. We use the following formulation to represent a GeMM:
\begin{equation}
    M\times X=Y
\end{equation}
where the $M$ is a model weight and $X$ is a matrix with each column as an activation vector. Usually, the model weight matrix is stored in a low-precision format such as int4 or fp8 and the activation matrix is stored in a higher precision format such as fp16 or fp32. When a datatype is as small as an int4, arithmetic operations such as multiplication may no longer need an actual functional unit in the processor, instead a simple $2^4\times 2^4$ multiplication table would be sufficient where each row and column would correspond to one of the $2^4$ possible values of each operand. This paper applies a similar idea to GeMMs with low-precision $M$ and arbitrary precision $X$ and $Y$. Specifically, we proposes \msgemm (Microsoft GeMM), a new technique in computing GeMMs with low-precision model weights using look-up tables such that the number of required arithmetic operations is $2.5\times$ lower than traditional GeMM computation.

\section{Background}
Transformer model architecture~\cite{transfomer} is widely used in industry and worked on in the research community. At the core, most operations in a transformer layer is a GeMM operation along with some non-linear operations such as layernorm or ReLU. The focus of this paper is on the GeMM operations as they take the majority of the end-to-end time. 

\begin{figure}
    \centering
    \includegraphics[width=.5 \textwidth]{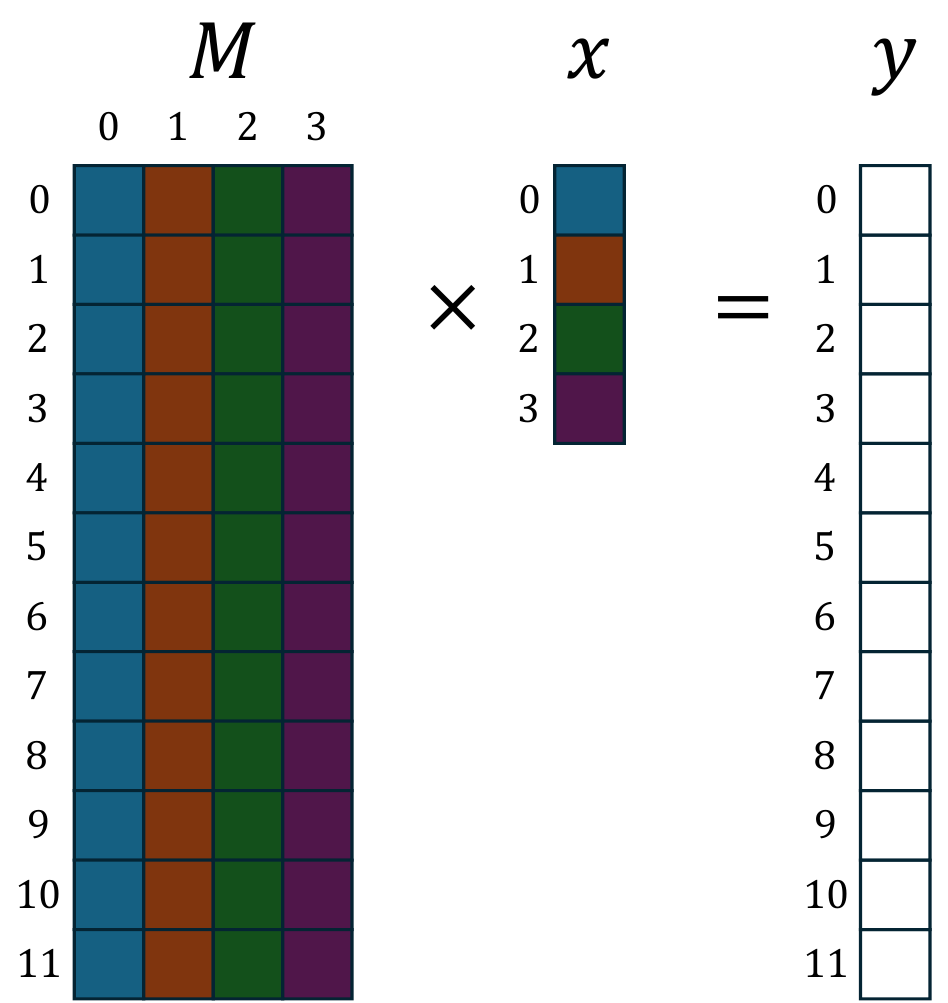}
    \caption{Multiplication of an MLP matrix $M$ by an activation vector $x$ with an output vector $y$.}
    \label{fig:mxy}
\end{figure}

A GeMM operation in a transformer layer is usually represented by $M\times X=Y$ where $M$ is the weight matrix, $X$ is the activation matrix (a batch of different activation vectors stacked as a matrix), and $Y$ is the output matrix (a batch of outputs stacked as a matrix). The computation of a GeMM is represented by $m\times k\times b$ where $m$ is the number of rows of $M$ and $Y$, $k$ is the number of columns of $M$ and the number of rows of $X$, and $b$ is the batch size which is the number of columns of $X$ and $Y$.
%As a running example in this paper, we will consider the first MLP GEMM size from GPT-3~\cite{gpt3} which is $4H\times H\times B$ where $H=12288$ is the hidden dimension size and $B$ is the batch size. For simplicity, we assume $B=1$ for now. Therefore, we express the computation by $M\times x=y$ where $M$ is the MLP matrix weight with size of $4H\times H$ and $x$ is the activation vector of size $H\times 1$ for the single batch. Figure 
As a running example in this paper, we will consider a GeMM with a computation size of $12\times 4\times 1$ and later we will generalize it to any GeMM size. Figure~\ref{fig:mxy} shows $M\times x=y$ for this example\footnote{We will use lower-case $x$ and $y$ for representing vectors and upper-case $X$ and $Y$ for matrices.}. Note that in this figure, all blue elements of $M$ ($\forall i: M(i,0)$) need to multiply with the blue element of $x$ ($x(0)$), red with red and so on. Since there are many blue elements of $M$ multiplied by the same blue element of $x$, one may wonder if the same multiplication is repeated too many times when $M$ has a low-precision representation.
%where $H=12288$ is the hidden dimension size and $B$ is the batch size. For simplicity, we assume $B=1$ for now. Therefore, we express the computation by $M\times x=y$ where $M$ is the MLP matrix weight with size of $4H\times H$ and $x$ is the activation vector of size $H\times 1$ for the single batch. Figure 

There are many proposed datatypes and precision for both matrix weights and activation vector including floating points and integers. The key contribution of this paper is that if either the matrix weights or activation vector has a low-precision datatype such as a 4-bit representation, the number of required operations can be significantly reduced. Therefore, for the rest of this paper, we will assume that the weight matrix is in 4-bit int (int4) and the activation vector is in any arbitrary datatype. Note that the exact datatypes and their arithmetic are irrelevant. We could have 8-bit floating point for weight matrix and 32-bit floating for activation vector and similar conclusions could be made.

Consider Figure~\ref{fig:mxy} where $M$ has $12$ rows and the datatypes are in int4. Given that int4 has only $2^4=16$ possible values and half of them are negative, at least there are two rows on the same column that have the same absolute values. Therefore, $\forall i: M(i,0)\cdot x(0)$, at least two of them are off by just a sign. This means that some of the multiplications can be avoided by looking up previously computed values. However, looking up a single multiplication is not going to help much as a GeMM operation requires both a multiplication and an addition. Therefore, the key question is whether we can also have the addition as a part of the look-up.

For now, we assume a GeMM operation is a matrix-vector multiplication represented by $M\times x=y$ and we will later generalize it later.

\section{\msgemm Algorithm}
\msgemm algorithm takes advantage of the low-precision datatype of matrices in AI GeMM operations to reduce down on the number of required arithmetic operations and it works in two phases: (1) producing a look-up table, (2) consuming the look-up table. Next we will explain each phase separately:

\subsection{Producing the Look-Up Table}
\label{sec:prod-lut}
The algorithm has a parameter denoted by $d$ which corresponds to the depth of accumulation into the look-up table. As a reminder, $M$ matrix is of size $m\times k$ and $x$ is of size $k\times 1$. Our look-up table $L$ is of size $\underbrace{2^4\times 2^4\times \cdots 2^4}_d\times \frac{k}{d}$ and defined as follows\footnote{We assume $k$ is divisible by $d$ in this paper.}:
\begin{align}
    &\forall p: 0000 \leq i_p \leq 1111, \forall j: 0\leq j < \frac{k}{d}: L(i_0, i_1, \cdots, i_{d-1},j) \\
    &=\sum_{r=0}^{d-1} b(i_r)\cdot x(j\cdot d + r) \label{f1}
\end{align}
where $b(i)$ is the $i^{th}$ possible value for the datatype of $M$ which in our example is int4 and therefore, there are $2^4$ possible values. One can think of $b$ as a function from a 4-bit representation to an int4 value: $b(0000)=0$, $b(0001)=1$, ..., $b(0111)=7$, $b(1000)=-8$, ..., $b(1111)=-1$.

\begin{figure}
    \centering
    \includegraphics[width=.9 \textwidth]{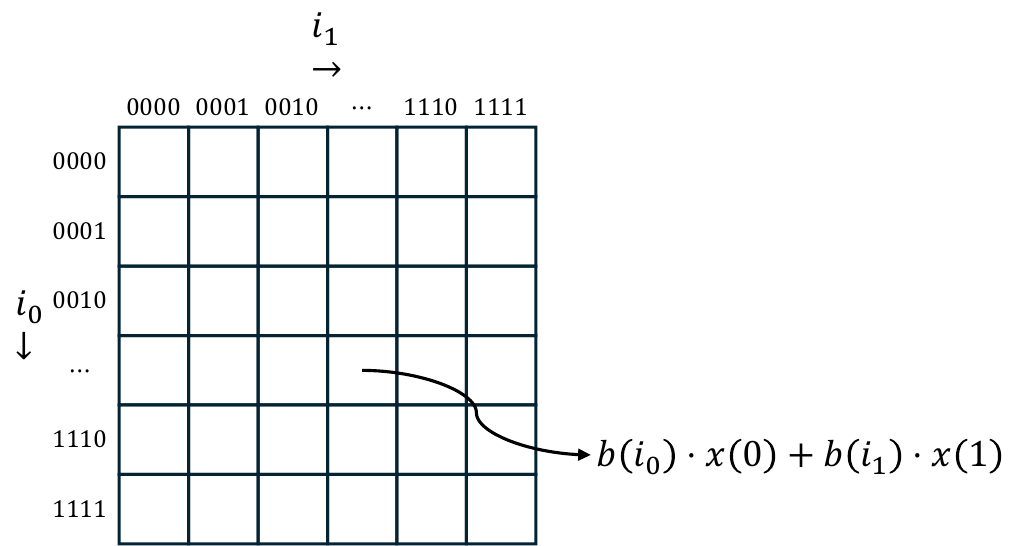}
    \caption{A 2D block of the look-up table $L$ for Figure~\ref{fig:mxy} with $d=2$. Note that $L$ is actually a 3D table and this is only showing a 2D block of it by fixing the last dimension.}
    \label{fig:lut}
\end{figure}

\paragraph{Example} Suppose that $k=4$ as shown in Figure~\ref{fig:mxy} and $d=2$. $L$ has a size of $2^4\times 2^4\times 2$. Let's focus on a 2D block of $L$ by fixing the last dimension i.e. $\forall i_0,i_1: L(i_0,i_1,0)$. Figure~\ref{fig:lut} shows this 2D block for $L$. Each element of this table stores the value of $b(i_0)\cdot x(0) + b(i_1)\cdot x(1)$.

As it can be implied, the computation of this look-up table is dependent on the activation vector $x$ and effectively, it constructs all possible linear combinations of $d$ int4s with $d$ consecutive elements of $x$ vector as coefficients. Therefore, one can use this table to find pre-computed values for $M\times x$ instead of computing it, specially when $M$ has many more rows than the number of elements of the look-up table.

\subsection{Consumer of Look-Up Table}
The second phase of the algorithm consists of consuming the look-up table built in Section~\ref{sec:prod-lut}. Let's consider Figure~\ref{fig:mxy} again and look-up table $L$ with $d=2$ already computed as shown in Figure~\ref{fig:lut}. Therefore, for every possible int4 values that may exist on blue and red columns of matrix $M$, we have already pre-computed the linear combination of them with $x(0)$ and $x(1)$, the blue and red rows of $x$. Similarly, all linear combination of $x(2)$ and $x(3)$ is already pre-computed. Therefore, each element of $y$ is simply a summation of two elements of the look-up table. For example, if assume that the first row of $M(0,:)=\{2,4,3,5\}$, then $y(0)=L(0010,0100,0)+L(0011,0101,1)$.

This can be extended to any general GeMM formulation as follows:
\begin{align}
    y(i) &= \sum_{j=0}^{k-1} M(i,j)\cdot x(j)=\sum_{j=0}^{\frac{k}{d}-1}\sum_{r=0}^{d-1}M(i,j\cdot d + r)\cdot x(j\cdot d + r) \label{f2:line0} \\
         &=\sum_{j=0}^{\frac{k}{d}-1}L\Big(\hat{b}\big(M(i,j\cdot d + 0)\big), \hat{b}\big(M(i,j\cdot d + 1)\big), \label{f2:line1} \\
         &\cdots, \hat{b}\big(M(i,j\cdot d + d-1)\big), j\Big)
\end{align}
where $\hat{b}$ is the inverse of $b$ described in Section~\ref{sec:prod-lut} i.e. it gives the 4-bit representation of a given value. For example, $\hat{b}(-1)=1111$. Equation~\ref{f2:line0} is straight forward and it is from the naive way of computing a GeMM. The inner summation looks similar to Equation~\ref{f1} where $L$ has the pre-computed all possible linear combination of int4 values with $\forall r: 0\leq r<d: x(j\cdot d + r)$. Therefore, Equation~\ref{f2:line1} is a direct application of the look-up table by finding the 4-bit representation of each element of $M$.

It should be clear how the proposed \msgemm algorithm uses the look-up table instead of computing each element directly. However, one may wonder if this is ever cost effective given that we have to pre-compute a large look-up table which will be discussed next.

\subsection{Datatypes with Shared Scales}
Custom datatypes where a block of elements of $M$ share a scale are common in AI workloads. For example, MSFP12~\cite{msfp} for matrix $M$, stores all elements in int4 as the mantissa of a floating point and several elements in a bounding-box share an 8-bit exponent so that each element has a 12-bit floating point representation. As an instance, a bounding-box around elements of a row of $M$ means that $\forall i: y(i)=q(i)\cdot \sum_j M(i,j)\cdot x(j)$ where $q$ is an array of a shared exponents for each row of $M$. One can think of $q(i)$s as any arbitrary scale in floating point as well. Note that this type of scaling is directly applicable to \msgemm since we can multiply each $y(i)$ by $q(i)$ after consuming $L$. However, the bounding-box could be arranged in a 2D shape where only a few elements of each row share a scale.

In general, \msgemm is still applicable if $r$ elements of a row of $M$ share a scale where  $r\geq d$ and preferably $r$ is a multiplication of $d$. This is because the scale factor can be applied after consuming the look-up table. However, if the bounding-box was, for example, around each column of $M$, \msgemm would not be applicable. Regardless of \msgemm, this format of bounding-box is inefficient for the hardware since computing each $y(i)$ would require $2\cdot k$ multiplications (one extra multiplication for the scale of elements of $M$) and $k-1$ addition. Luckily, if the bounding-box has multiple elements of a row, this extra multiplication can be factored out and implementing it becomes more efficient in the hardware. This also aligns with \msgemm perfectly as it only requires $d$ elements of each row to share a scale when there is one. As we will show in Section~\ref{sec:results}, $d\leq4$ for practical cases. Therefore, \msgemm is applicable for when at least $d$ elements of a row share a scale which we expect to be the case in most scenarios. 

\section{Complexity}
\label{sec:complexity}
In this section, we will calculate the complexity of each phase of the algorithm. As before, let's assume that $M$ is of size $m\times k$. Given that the depth of $L$ is $d$, it has  $\underbrace{2^4\times 2^4\times \cdots 2^4}_d\times \frac{k}{d}=2^{4\cdot d}\times \frac{k}{d}$ elements as explained in Section~\ref{sec:prod-lut}. Each element of $L$ is a linear combination of $d$ elements which means it costs $d-1$ additions and $d$ multiplications (the result of the first multiplication does not need an add). Let's round that up to $d$ fused multiplication-add operations. Therefore, the computation complexity of $L$ is:
\begin{equation}
    C(L) = 2^{4\cdot d}\cdot \frac{k}{d}\cdot d=2^{4\cdot d}\cdot k
    \label{eq:lut-com}
\end{equation}
Number of memory access for $L$ is just the access of vector $x$. The 4 bits for the linear combination of each element is instead constructed as opposed to a memory access. Therefore, the amount of memory access for $L$ is:
\begin{equation}
    M(L) = k
    \label{eq:lut-mem}
\end{equation}

The complexity of the second phase of the algorithm can be calculated by considering Equation~\ref{f2:line1} which requires adding $\frac{k}{d}$ elements from $L$ table to compute a single element of $y$. Considering that $y$ has $m$ rows, the computation complexity of $y$ by consuming $L$ is:
\begin{equation}
    C(y) = (\frac{k}{d}-1)\cdot m
    \label{eq:y-com}
\end{equation}
Note that there is no cost associated with $\hat{b}$ function from Equation~\ref{f2:line1}. We assumem that $M$ is stored in a row-major format and $d$ consecutive elements of $M$ in int4 representation can be thought as $d$ concatenated int4 together to form a int4$d$ which can be used directly to dereference the first $d$ dimension of $L$. Therefore, there is no computation required to indexing into $L$.

The memory accesses required for the second phase of computation is simply the number of elements of $M$ which is $m\times k$. We assume that $L$ computed in phase 1 is kept in cache. If $L$ is too large to be kept in cache, we can pipeline phase 1 and 2 by partially computing $L$ and fixing the last dimension. Therefore, the memory access required for the second phase of \msgemm is:
\begin{equation}
    M(y) = m\cdot k
    \label{eq:y-mem}
\end{equation}

The total computation and memory access of \msgemm for both phases are:
\begin{align}
    C(\msgemm) &= C(L)+C(y)=2^{4\cdot d}\cdot k+(\frac{k}{d}-1)\cdot m \\
    M(\msgemm) &= M(L)+M(y)=k+ m\cdot k
    \label{eq:cost}
\end{align}

Note that a naive GeMM computation requires $(k-1)\cdot m$ fused multiplication-add operations and $m$ multiplication (the first multiplication does not need an addition). Let's round that up to $k\cdot m$ fuse multiplication add operations\footnote{In most modern processors, addition, multiplication, and fused multiplication-add cost the same.}. The memory access of a naive GeMM computation is $k+m\cdot k$ for each element of $M$ and $x$. The number of memory accesses of a naive GeMM and \msgemm are identical. The computation, however, is quite different. For \msgemm, $C(y)$ is nearly $d$ times smaller than the naive GeMM computation. But, $C(L)$ is the obvious overhead in this algorithm and we need to consider when this overhead is manageable. The exponential growth rate of $2^{4\cdot d}$ asks for a small $d$ but note that $C(L)$ is independent of $m$, the number of rows of $M$. Therefore, as $m$ grows, the overhead of $L$ minimizes. The key question is whether GeMM operations in modern LLMs have large enough $m$ for \msgemm to be efficient. We will study this in Section~\ref{sec:results}.

\subsection{Optimization}
There are many obvious optimizations that can be applied to the computation of the look-up. The most obvious one is utilizing the linear property of linear combination of elements of $L$: $\sum_r q\cdot b(i_r)\cdot x(r)=q\cdot(\sum_r b(i_r)\cdot x(r))$ for any $q$. For example, for $q=-1$, we can {\em almost} calculate $L$ for only the non-negative values of $b(i_0)$. But this comes with a lot of caveats. For example, he datatype may not allow negative values (uint4) or even if they do (int4), the reduction in computational complexity of $L$ is incredibly complicated. For example, int4 has $8$ negative values, $7$ positive values and a $0$ and computing with only non-negative values of $b(i_0)$ is not enough. Other datatypes such as fp8 has special set of bits reserved for inf or nan and etc which would be hard to capture all of them for optimizing $L$. Therefore, we consider $L$ for general cases.

\subsection{Generalization}
\label{sec:gen}
From the beginning, we assumed that $x$ is an activation vector which corresponds to a batch size of $1$. In general, GeMM operations in AI workloads run with larger batch sizes. Luckily, all of the computational complexity calculated in this section scales linearly with the batch size and the relative benefit of \msgemm over naive GeMM computation remain unchanged. For the memory access, both naive GeMM and \msgemm will have $k\cdot b+m\cdot k$ accesses and they will stay identical. Therefore, for the rest of this paper, we assume that the cost of the GeMM of size $m\times k\times b$ is
\begin{align}
    C(\msgemm) &= \Big(2^{4\cdot d}\cdot k+(\frac{k}{d}-1)\cdot m\Big)\cdot b \\
    C(GeMM) &= m\cdot k \cdot b
    \label{eq:cost:both}
\end{align}

\section{Theoretical Evaluation}
\label{sec:results}
In this section, we will compare the cost of \msgemm against the cost of naive GeMM computation and evaluate when \msgemm has an advantage for AI GeMM operations.

The speedup of \msgemm over naive GeMM can be calculate using Equation~\ref{eq:cost:both} in Section~\ref{sec:gen}:
\begin{align}
    \frac{C(GeMM)}{C(\msgemm)} = \frac{m\cdot k \cdot b}{\Big(2^{4\cdot d}\cdot k+(\frac{k}{d}-1)\cdot m\Big)\cdot b}=\frac{m\cdot k}{2^{4\cdot d}\cdot k+(\frac{k}{d}-1)\cdot m}
    \label{eq:speedup}
\end{align}

Now, let's pick some real-world GeMM operations from AI workloads. GPT-3~\cite{gpt3} models architecture consists of two GeMM operations from the MLP module of the transformer layers, $MLP_1$ and $MLP_2$, which are of size $12288\times 49152\times b$ and $49152\times 12288\times b$. Therefore, the speedup number for a given $d$ would be:
\begin{align}
    speedup(MLP_1) &=\frac{m\cdot k}{2^{4\cdot d}\cdot k+(\frac{k}{d}-1)\cdot m}\\
                   &=\frac{12288\times 49152}{2^{4\cdot d}\cdot 49152+(\frac{49152}{d}-1)\cdot 12288} \\
                   &= \frac{49152}{2^{4\cdot d}\cdot 4+(\frac{49152}{d}-1)} \\
    speedup(MLP_2) &=\frac{m\cdot k}{2^{4\cdot d}\cdot k+(\frac{k}{d}-1)\cdot m}\\
                   &=\frac{49152\times 12288}{2^{4\cdot d}\cdot 12288+(\frac{12288}{d}-1)\cdot 49152} \\
                   &= \frac{49152}{2^{4\cdot d}+(\frac{12288}{d}-1)\cdot 4}
    \label{eq:speedup}
\end{align}

\begin{figure}[htbp]
\centering
\input{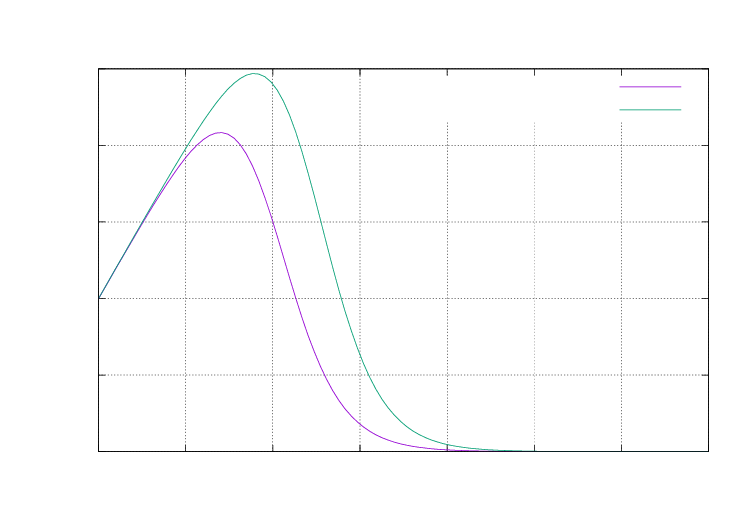}
\caption{Comparing the performance of the 2-phase algorithm against the naive computation of GeMM.}
\label{fig:speedup}
\end{figure}

Figure~\ref{fig:speedup} shows the speedup of the proposed algorithm for varying $d$, the depth of the look-up table. Because of the exponential cost growth of $L$, $d$ cannot be larger than $4$. However, a value of $3$ shows a sweet-spot of $\approx 2.5\times$ speedup for \msgemm for both MLP GeMM operations. Another observation in this plot is that the larger the number of rows ($m$) the better this algorithm works and that is because the cost the look-up table is independent of $m$.

\section{Implementation: Proposal for Next-Gen GPUs}
GPUs are the common backend target for running GeMM computation and they have been heavily optimized for naive GeMM computation by utilizing Tensor Cores~\cite{tensorcore}. A100 GPUs have an impressive 312 TFLOPS for fp16 GeMMs with Tensor Cores. However, the CUDA cores of A100s (normal computational cores) are limited to 19.5 TFLOPS. This means that the naive GeMM computation can run at 312 TFLOPS. Similarly, the first phase of \msgemm can run at 312 TFLOPs. However, the second phase of this algorithm requires adding elements of the the look-up table where Tensor Cores cannot be utilized. Therefore, the second phase limited to 19.5 TFLOPS. This is a limiting factor of current GPUs which do not allow \msgemm to be implemented efficiently.

This shortcoming can be addressed by adding special CUDA cores in GPUs which allow for adding elements of a look-up table with a similar performance of Tensor Cores.

\section{Conclusion}
This paper proposes \msgemm which utilizes the low-precision property of GeMM operation in AI workloads to reduce down the total required number of multiplication and add. We show that by using a look-up table with depth of $3$, modern AI GeMMs can be heavily optimized by $\approx 2.5\times$. However, this type of optimization requires special support from the hardware which current generation of GPUs do not have.

\end{document}